\documentclass[aps,pra,twocolumn,showpacs]{revtex4-1}
\usepackage{amsbsy,latexsym,amsmath}
\usepackage{amsfonts}
\usepackage{amssymb}
\usepackage[mathscr]{eucal}
\usepackage{epsfig,graphics,graphicx}

\newcommand{\id}{\openone}

\newcommand{\tr}{{\rm tr}\,}

\newcommand{\ket}[1]{\left|{#1}\right\rangle}

\newcommand{\braket}[2]{\langle{#1}|{#2}\rangle}
\newcommand{\ketbrad}[1]{\left|{#1}\rangle\!\langle{#1}\right|}

\newcommand\xyZ[3]{\mbox{\tiny $\kern-.3em(\!#1\kern-.2em #2\!)\kern-.14em #3\!$}}
\newcommand\Xyz[3]{\mbox{\tiny $\kern-.3em #1\kern-.14em(\! #2\kern-.14em #3\kern-.14em)\!$}}

\begin{document}
\title{Programmable discrimination with an error margin}

\author{G.~Sent\'{i}s$^{1}$, E.~Bagan$^{1,2}$, J.~Calsamiglia$^{1}$ and R.~Mu\~{n}oz-Tapia$^{1}$}

\affiliation{$^{1}$F\'isica Te\`orica: Informaci\'o i Fen\`omens Qu\`antics, Universitat
Aut\`{o}noma de Barcelona, 08193 Bellaterra (Barcelona), Spain\\ 
$^{2}$Centre for Quantum Technologies, National University of Singapore,
3 Science Drive 2, Singapore 117543, Singapore}

\begin{abstract}


The problem of optimally discriminating between two completely unknown qubit states 
 is generalized by allowing an error margin. It is visualized as a device---the programmable discriminator---with one data and two program ports, each fed with a number of identically prepared qubits---the data and the programs. The device aims at correctly identifying the data state with one of the two program states. This scheme has the unambiguous and the minimum error schemes as extremal cases, when the error margin is set to zero or it is sufficiently large, respectively. Analytical results are given in the two situations where the margin is imposed on the average error probability---weak condition---or it is imposed separately on the two probabilities of assigning the state of the data to the wrong program---strong condition. It is a general feature of our scheme that the success probability rises sharply as soon as a small error margin is allowed, thus providing a significant gain over the unambiguous scheme while still having high confidence results.

\end{abstract}

\maketitle
\section{Introduction}

Quantum state discrimination is one of the most basic yet fundamental tasks in quantum information~\cite{helstrom}. In its simplest form, it consists in a
protocol that tells in which out of two given states a quantum system was prepared. This is a primitive of great practical interest that has been 
investigated from many perspectives and for which many key results have been obtained. Theoretical results also abound in the literature, e.g., state discrimination provides an operational distance between any two states~\cite{chernoff} based on the degree of difficulty of telling one from the other. It has also been shown that for multiple copies of pure states there exist individual adaptive measurements on each copy that provide exactly the  same discrimination power as the 
optimal (global) measurement strategy~\cite{bayes-local}. This is however not so for mixed states, and there is numerical evidence that even the corresponding asymptotic exponential error rates are different in this case~\cite{chernoff-locc,australians-1}.


Generically, a discrimination protocol, to which we will refer throughout the paper as  {\it device}, {\it machine} or more explicitly as {\em discriminator},
is not universal but
specifically designed for each given 
pair of possible states.
A significant conceptual twist on discrimination was introduced in~\cite{dusek,bergou-hillery}, where devices that work
for arbitrary pairs of states were considered. 
%
%
These machines have two program ports through which multiple copies of the unknown quantum states are loaded (``the programs,'' for short). 
Multiple copies of a third state (guaranteed to coincide with one of the states loaded through the program ports) are fed into the data port of the machine. This so-called {\em programmable discriminator} is designed to report whether the state of the data is that of the first program, or whether it is that of the second program.
%
%
The discrimination 
protocol exploits the difference between the permutation symmetry of the global state of the three ports in the two alternatives.
These machines work  for discrete~\cite{bergou-hillery,sentis} as well as for continuous variable systems~\cite{sedlak}.
Programmable discriminators can be regarded as machine-learning devices. It has recently been shown that, 
in some settings,
optimal performance can be attained with a suitable measurement on the two programs followed by a measurement on the data, where only classical communication between the two separate measurements is required.
Not only does this mean an important saving of resources, as conventional memory suffices to store the (classical) output of the first measurement, but also that
programmable discriminators  can be reused and still exhibit optimal performance without having to reload the program ports~\cite{us-rep}.
Interestingly, programmable discrimination is also formally equivalent to a change-point problem \cite{masahito}. Let us assume that a source produces
states of an unknown type
and that either at time $t_1$ or at time $t_2$ the same source starts producing states of a different type. The change-point problem 
consists in 
identifying whether the time at which the change occurs is $t_1$~or~$t_2$. 

In most of the literature so far either the minimum-error or the unambiguous discrimination scheme is considered. In the former, the discriminator always
produces a conclusive answer about the identity of the input state, but sometimes this answer is  wrong. In the unambiguous scheme no error  is allowed, that is, the input state must be correctly identified with certainty. This can only happen at the expense of producing
some inconclusive answers or, in other words, the machine  sometimes must {\em abstain}~\cite{gendra} from giving an answer. 
In both cases optimality means that the machine attains maximum success probability.  It is clear that, if we relax the unambiguous scheme by tolerating
some error rate, we can increase the success probability. Likewise, by allowing 
some rate of inconclusive answers in the minimum-error scheme, 
we can also increase the reliability of the answers. Hence by introducing an error margin 
we can unify minimum-error and unambiguous discrimination. Both become extremal points of the unified discrimination with error margin scheme~\cite{hayashi1,hayashi,q-discr}. Interpolating between these two extemal cases may have practical interest in some situations.

In this paper we combine the two concepts above and analyze the optimal performance of a qubit multiple-copy programmable machine when an error rate is allowed. We will show that by relaxing the zero error condition slightly the resulting scheme provides an important enhancement in performance over the widely used unambiguous scheme. We will first review the standard problem, when the
states between which we wish to discriminate are known.  For the sake of self-containedness, we will rederive the success probability for a given error margin in both the so-called weak and strong senses.
We will then present our results for  programmable devices and  obtain the analytical expression of the success probability as a function of the error margins. We will discuss our results in a separate section and will end the paper by stating our conclusions.

\section{Discrimination with error margins}

Consider two pure nonorthogonal states 
$\rho_1=\ketbrad{\psi_1}$, $\rho_2=\ketbrad{\psi_2}$ 
as hypotheses of a standard two-state discrimination problem, where for simplicity we assign equal prior probabilities to each state. The discrimination with an error margin protocol can be thought of as a 
generalized  measurement on the system, described mathematically by a positive operator-valued measure (POVM) with three elements $\mathcal{E}=\{E_1,E_2,E_0\}$, where the operator $E_1$ ($E_2$) is associated to the statement ``the measured state is $\rho_1$ ($\rho_2$),'' whereas  $E_0$ is associated to the inconclusive answer or abstention. The overall success, error, and inconclusive probabilities are
$
P_{\rm s} = \frac{1}{2} \left[ \tr (E_1 \rho_1) + \tr(E_2 \rho_2) \right]$,
$
P_{\rm e} = \frac{1}{2} \left[ \tr (E_2 \rho_1) + \tr(E_1 \rho_2) \right] $,
and $Q = \frac{1}{2} \left[ \tr (E_0 \rho_1) + \tr(E_0 \rho_2) \right] $,
respectively. The relation $P_{\rm s}+P_{\rm e}+Q=1$ is guaranteed by the POVM condition~$E_0+E_1+E_2=\openone$.
The optimal discrimination  with  an error  margin  protocol is obtained by maximizing 
the success probability $P_{\rm s}$ over any possible POVM~$\mathcal{E}$ that satisfies that certain errors occur with a probability not exceeding the given margin. Generically,  these conditions imply a nonvanishing value of  the inconclusive  probability $Q$.

In this paper, we consider two error margin conditions: {\em weak} and {\em strong}. The weak condition states that the \emph{average} error probability cannot exceed a margin,~i.e.,
\begin{equation}\label{weakgeneral}
P_{\rm e} = \frac{1}{2} \left[ \tr (E_2 \rho_1) + \tr(E_1 \rho_2) \right] \le  r \,.
\end{equation}
The strong condition imposes a margin on the probabilities of misidentifying \emph{each} possible state, i.e.,
\begin{eqnarray}
p(\rho_2|E_1)&=&\frac{\tr (E_1\rho_2)}{\tr (E_1\rho_1)+\tr (E_1\rho_2)} \le r \label{strong1}\, ,\\[.5em]
p(\rho_1|E_2)&=&\frac{\tr (E_2\rho_1)}{\tr (E_2\rho_1)+\tr (E_2\rho_2)} \le r \label{strong2}\, ,
\end{eqnarray}
where $p(\rho_2|E_1)$ and $p(\rho_1|E_2)$ are the probabilities that the state identified as $\rho_1$ is actually $\rho_2$ and the other way around, respectively. 
The strong condition  is obviously 
more restrictive, 
as it sets a margin on both types of errors separately. However, as we will see, the two conditions are directly related: the strong one just corresponds to the weak one with a tighter error margin~\cite{hayashi}.
%
Note that both error margin schemes have the unambiguous (when~$r=0$) and the minimum-error schemes (when~$r$ is large enough) as extremal cases. 
We will denote by $r_c$ the critical margin above which the success probability does not increase and thus coincides with that of (the unrestricted) minimum-error discrimination.

\begin{figure}
\includegraphics[scale=.25]{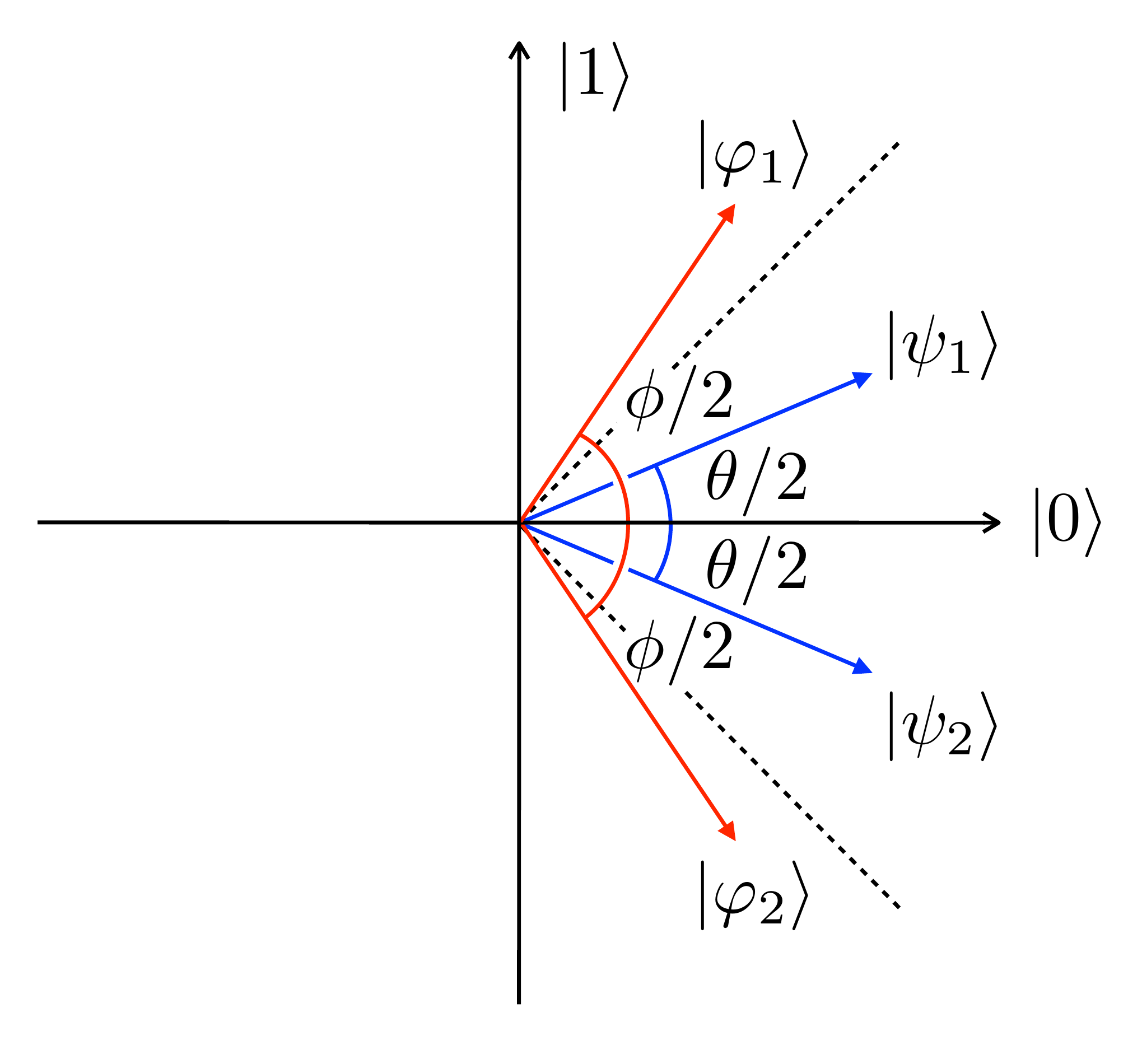}
\caption{(Color online) Parametrization of the states $|\psi_1\rangle$, $|\psi_2\rangle$, $|\varphi_1\rangle$, and $|\varphi_2\rangle$ as in Eqs.~(\ref{the_states}) and~(\ref{the_POVM}).}\label{fig3}
\end{figure}

For the weak condition, it is straightforward to obtain the maximum success probability
by taking into account that the corresponding error probability must saturate the margin condition~\eqref{weakgeneral} for~$r \le r_c$, namely, $P_{\rm e}=r$. Furthermore, the symmetry of the problem dictates that $\tr (E_1 \rho_1)=\tr (E_2 \rho_2)=P_{\rm s}$ and~\mbox{$\tr (E_1 \rho_2)=\tr (E_2 \rho_1)=P_{\rm e}$}. Without loss of generality (see Fig.~\ref{fig3}), we can write the input states as
%
\begin{equation}
\ket{\psi_i}=\cos{\theta\over2}\ket{0}-(-1)^{i}\sin{\theta\over2}\ket{1}\,, \quad i=1,2\,,
\label{the_states}
\end{equation}
where $0\le \theta\le \pi/2$,
and the POVM elements as~$E_i=\mu \ketbrad{\varphi_i}$ for $i=1,2$, with
\begin{equation}
\ket{\varphi_i}=\cos{\phi\over2}\ket{0}-(-1)^{i}\sin{\phi\over2}\ket{1}, \quad
{\pi\over2}\le\phi\le \pi.
\label{the_POVM}
\end{equation}
The POVM condition implies $E_0=\openone -E_1-E_2$, and the optimal value of~$\mu$ is fixed by the extremal value of the 
inequality $E_0\ge 0$. One obtains $\mu=1/(1-\cos\phi)\le1$ and finally the symmetry conditions fix~$\phi$ to be
%
\begin{equation}\label{optimal-phi}
\tan\frac{\phi}{2}=
\begin{cases}
\displaystyle\frac{\sqrt{1+c}}{\sqrt{1-c}+2\sqrt{r}} & \text{if} \quad 0\le r\le r_c \,,  \\
1 & \text{if} \quad r_c\le r\le 1 \, ,
\end{cases}
\end{equation}
where  $c=|\braket{\psi_1}{\psi_2}| = \cos \theta$ is the overlap of the states~$\ket{\psi_1}$ and~$\ket{\psi_2}$.
Notice that in the unambiguous limit, $r=0$,  the POVM elements~$E_1$ and~$E_2$ are orthogonal
to the states~$\ket{\psi_2}$ and~$\ket{\psi_1}$, respectively.
%
In the other extreme case, when the error margin coincides with, or is larger than, the minimum error,~\mbox{$r\ge r_c$}, one has $E_0=0$ (no abstention) and $E_1$ becomes orthogonal to $E_2$, i.e.,~$\phi=\pi/2$. In this range the measurement becomes of von Neumann type and the first case in~(\ref{optimal-phi}) implies 
\begin{equation}\label{critical-r}
r_c = \frac{1}{2} \left( 1-\sqrt{1-c^{2}} \right) .
\end{equation}
Taking into account Eq.~\eqref{optimal-phi}, the optimal success probability reads 
%
\begin{equation}\label{weak}
P_{\rm s}^{W}(r) =
\begin{cases}
\left( \sqrt{r} + \sqrt{1-c}\, \right)^{2} & \text{if} \quad 0\le r\le r_c \, , \\
\frac{1}{2} \left( 1+\sqrt{1-c^{2}} \,\right) & \text{if} \quad r_c\le r\le 1 \, .
\end{cases}
\end{equation}
%
%
%
This result was derived in~\cite{hayashi1} and its generalization to arbitrary prior probabilities in~\cite{hayashi} (also in~\cite{q-discr}, by fixing an inconclusive rate $Q$ instead of an error margin).
Note that the POVM $\mathcal{E}$ is fully determined by the angle $\phi$, which in turn is fully determined by the margin $r$ through Eq.~\eqref{optimal-phi}.

The optimal success probability under the strong condition can be obtained along the same lines of the weak case, but it will prove more convenient to use the connection between both conditions to derive it directly from~\eqref{weak}.
Let us denote by  $r^S$  ($r^W$) the error margin of the strong (weak) condition. 
From the symmetry of the problem, 
Eqs.~\eqref{strong1} and~\eqref{strong2} can be written in the form of a weak condition with  a margin $r^W$ as
%
%
\begin{equation}
P_{\rm e} \le r^S(P_{\rm e}+P_{\rm s}) \equiv r^W .
\end{equation}
%
Hence, if $\mathcal{E}$ is the optimal POVM for a strong margin $r^S$, it is also optimal for the weak margin $r^W$, where $P_{\rm e}=r^W$ and $P_{\rm s}=P_{\rm s}^W(r^W)$ is given by Eq.~\eqref{weak}.
In terms of the success probability, the relation between $r^W$ and $r^S$ reads
\begin{equation}\label{mwms1}
 r^S=\frac{r^W}{P_{\rm s}^W(r^W)+r^W}\, .
\end{equation}
%
%
By solving for $r^W$ and substituting into Eq.~\eqref{weak} one derives the success probability for a given $r^S$, which we denote by $P_{\rm s}^S(r^S)$. For
the function $P_{\rm s}^S$ one readily obtains
\begin{equation}\label{strong}
P_{\rm s}^{S}(r)\! =\!
\begin{cases}
\displaystyle\!\left( \!\frac{\sqrt{1\!-\!r}}{\sqrt{r}-\!\sqrt{1\!-\!r}} \right)^{2}\!(1\!-\!c) &\!\! \text{if} \quad 0\le r\le r_c \,, \\
\!\frac{1}{2} \!\left( 1\!+\!\sqrt{1\!-\!c^{2}} \,\right) &\!\!  \text{if} \quad r_c\le r\le 1 \,,
\end{cases}
\end{equation}
%
in agreement with~\cite{hayashi1}. Note that the critical margin is the same for both the weak and the strong conditions, i.e., $r_c^W=r_c^S=r_c$. 
Indeed, beyond the critical point inconclusive results are excluded by optimality ($Q=0$ and $P_{\rm s}+P_{\rm e}=1$) and thus there is no difference between the two types of conditions. As in the weak case, there is a correspondence between the angle $\phi$ and $r^S$; thus $\mathcal{E}$ can also be parametrized in terms of the strong margin:
%
\begin{equation}\label{phi-strong}
\tan\frac{\phi}{2}\!=\!
\begin{cases}
\displaystyle
\frac{\sqrt{1\!-\!r^S}-\!\sqrt{r^S}}{\sqrt{1\!-\!r^S}+\!\sqrt{r^S}}\,{\sqrt{1\!+\!c}\over\sqrt{1\!-\!c}}
& \text{if} \quad 0\le r\le r_c \, , \\[1em]
1 & \text{if} \quad r_c\le r\le 1 \, .
\end{cases}
\end{equation}
Note that 
an ambiguity arises for $c=1$, as $\phi=\pi$ and then $E_1$ and $E_2$ become proportional to one another, independently of the value of $r^S$.
Note also that \mbox{for $r^S=0$} and~$r^S=r_c$ the values of $\phi$ for both, weak and strong conditions, coincide.
\section{Programmable discrimination}

Let us elaborate on the definition of a programmable discriminator given in the Introduction. It is a device capable of identifying the state of a system (a qubit in our case) that is guaranteed to be prepared in one of two possible {\em unknown} pure states, say~$\{\ket{\psi_1},\ket{\psi_2}\}$.
By unknown we mean that we lack all the information about their preparation. Instead, we assume that we are supplied with~$n$ copies of each of them,
which can be fed into the device through two program ports labeled $A$ and~$C$ for~$\ket{\psi_1}$ and~$\ket{\psi_2}$, respectively. In addition, a third port~$B$ is loaded with~$n'$ copies of the state to be identified. A programmable discriminator is assumed to be a universal device and it should thus work for any pair of states~$\{|\psi_1\rangle,|\psi_2\rangle\}$. 
To make this paper self-contained, in this section
we review the state of the art of this discrimination problem. A more general and detailed analysis can be found in~\cite{sentis}. 

A programmable discriminator is defined by a universal 
POVM with three elements~$\mathcal{E}=\{E_1,E_2,E_0\}$. The operator $E_1$ ($E_2$) corresponds to the machine assigning the label 1 (2) to the copies in $B$, meaning that 
their state is identical to that of the copies in~$A$~($C$).
Once again, the third operator, $E_0$, is associated to an inconclusive result.
The optimal $\mathcal{E}$ is that which maximizes the averaged probability of success,
$
P_{\rm s} = \int d\psi_1 d\psi_2 P_{\rm s}(\psi_1,\psi_2)  \, ,
$
where~$P_{\rm s}(\psi_1,\psi_2)$ is the success probability for a given pair of states $\{\ket{\psi_1},\ket{\psi_2}\}$ and the average is taken over all possible pairs.
Since $\mathcal{E}$ is state independent, $P_{\rm s}$ can be recast as the success probability  of discrimination between the two effective global states (of the three-partite port system $ABC$) when the state in $B$ is either~$\ket{\psi_1}$ or~$\ket{\psi_2}$. These effective states are given by the averages
%
\begin{eqnarray}
\sigma_1&=&\int d\psi_1 d\psi_2 [\psi_1^{\otimes n}]_A [\psi_1^{\otimes n'}]_B [\psi_2^{\otimes n}]_C \, ,\nonumber \\
\sigma_2&=&\int d\psi_1 d\psi_2 [\psi_1^{\otimes n}]_A [\psi_2^{\otimes n'}]_B [\psi_2^{\otimes n}]_C\, ,
\label{average states}
\end{eqnarray}
respectively, where the notation $[\,\cdot\,]$ stands for $\ketbrad{\,\cdot\,}$. The integrals can be easily computed using the Schur lemma (see~\cite{sentis}) and one obtains
\begin{eqnarray}\label{sigmas}
		\sigma_1 &=&    \frac{1}{d_{AB} d_C} \openone_{AB} \otimes \openone_{C} 
\end{eqnarray}
and the analogous expression for $\sigma_2$ where the labels $A$ and $C$ are exchanged. Here
$\openone_{X}$ ($\openone_{XY}$) is the projector onto the completely symmetric subspace  of~$\mathcal{H}_X$ ($\mathcal{H}_X\otimes \mathcal{H}_Y$) and 
$d_{X}=\tr\openone_{X}$ ($d_{XY}=\tr\openone_{XY}$) is its dimension. In our case we have $d_A=d_C=n+1$ and $d_{AB}=d_{BC}=n+n'+1$. The states $\sigma_1$ and $\sigma_2$ are diagonal in the angular momentum basis $\{\ket{j\,m}\}$, but extra labels are needed to specify how the various subsystems~$A$, $B$, and $C$ are coupled to each other. In particular, we use the basis $\ket{(j_{\!A} j_{\!B}\!) j_{\!A\!B} j_C;\!jm}$ to diagonalize~$\sigma_1$ and~$\ket{j_{\!A}(j_{\!B}j_C\!)j_{\!BC};\!jm}$ to diagonalize~$\sigma_2$, where $j_A=j_C=n/2$, $j_B=n'/2$ and $j_{AB}=j_{BC}=(n+n')/2$. 
The diagonal form of $\sigma_1$ is
\begin{eqnarray}
\sigma_1&=& \frac{1}{d_{AB} d_C}\sum_{j=n'/2}^{n'/2+n} \sum_{m=-j}^{j} [(j_Aj_B)j_{AB}j_C;j m]\,, 
\end{eqnarray}
and the analogous form of~$\sigma_2$ is obtained by coupling $j_B$ and $j_C$ instead of $j_A$ and $j_B$.
The key property of the angular momentum basis is that it satisfies the orthogonality relation
\begin{equation}\label{overlap}
\braket{(j_{\!A}j_{\!B}\!)j_{\!A\!B} j_{C};\!j m}{j_{\!A}(j_{\!B}j_C\!)j_{\!BC} ;\!j' m'}\! =\!c_j\delta_{j j'} \delta_{m m'}\,,
\end{equation}
where the overlaps $c_j$ can be obtained from the Wigner~$6j$ symbols~\cite{edmonds} [see Eq.~(\ref{the_c_alpha}) below]. Bases obeying an orthogonality relation of the form~\eqref{overlap} exist for any two subspaces and are known as Jordan bases \cite{bergou-feldman-hillery}. Since a state of the first basis has nonzero overlap with only one element of the second basis, the problem of discriminating $\sigma_1$ from $\sigma_2$ can be cast as pure state discrimination in each Jordan subspace, which we label by $j$ (note that the overlaps $c_j$ do not depend on the magnetic number~$m$). Hence, the optimal POVM can be chosen to be of the form $\mathcal{E}=\bigoplus_j \mathcal{E}_j$, where each $\mathcal{E}_j$ is itself a POVM acting on the subspace ${\cal H}_j$ of total angular momentum~$j$, and the total success probability is simply the sum of all the contributions. The success probability for both, the unambiguous ($P_{\rm e}=0$) 
and 
the minimum-error ($Q=0$) schemes, are given respectively~by~\cite{sentis} 
%
\begin{eqnarray}
P_{\rm s}^{\rm UA} &=& \frac{n n'}{(n+1)(n'+2)} \, , \label{optUA}\\
P_{\rm s}^{\rm ME} &=&\frac{1}{2} +\frac{1}{2}\sum_{k=0}^n \frac{n'+2k+1}{(n+1)(n+n'+1)}\nonumber\\
&\times & \sqrt{1-\left[\frac{(n'+k)!n!}{(n'+n)!k!}\right]^2}  \, ,\label{optME}
\end{eqnarray}
where equal prior probabilities are assumed.

\section{Error margins in programmable discrimination}

In this section, we generalize programmable discrimination by allowing an error margin.
To ease the notation, 
rather than labeling the various subspaces ${\cal H}_j$ by their total angular momentum~$j$, we will simply enumerate them by natural numbers, $\alpha=1,2,\dots,n+1$, and sort them by increasing value of~$j$. Hence $j=\alpha+n'/2-1$. 
With a slight abuse of notation, we will accordingly write ${\cal H}_\alpha$ and enumerate the corresponding
POVMs and overlaps as ${\cal E}_\alpha$ and $c_{\alpha}$, respectively,  where one has~\cite{sentis}
\begin{equation}
c_\alpha={\begin{pmatrix} n'+\alpha-1\\ n' \end{pmatrix}\over\begin{pmatrix}n+n'\\ n'\end{pmatrix}}
\label{the_c_alpha} .
\end{equation}
%

A direct consequence of the block structure of the averaged states and $\mathcal{E}$ is that the overall success probability of a programmable discriminator can be expressed as
\begin{eqnarray}
P_{\rm s}\! &=&\!\sum_{\alpha=1}^{n+1} p_\alpha P_{{\rm s},\alpha} \, , \label{weakj} \\
p_\alpha \! &=&\! \tr(\sigma_i \id_\alpha)\!=\!\frac{2\alpha+n'-1}{(n+1)(n+n'+1)}\,, \quad i=1,2\,,
\end{eqnarray}
where $P_{{\rm s},\alpha}$ is the success probability of discrimination in the subspace ${\cal H}_\alpha$ and $p_\alpha$ is the probability of $\sigma_1$ and~$\sigma_2$ projecting onto that subspace 
upon performing the measurement $\{\id_\alpha\}$. Likewise, $P_{\rm e}$ and $Q$ can be expressed as a convex combination of the form~(\ref{weakj}). 

\subsection{Weak error margin}

Let us start by considering the weak condition. 
If~we denote the error margin by $R$, the weak condition reads~$P_{\rm e}\le R$.
%
According to the previous paragraph, the optimal strategy and the corresponding success probability $P_{\rm s}$ are defined through the maximization problem
\begin{equation}\label{max_probl}
P_{\rm s}\!=\!\max_{\cal{E}} \sum_{\alpha=1}^{n+1} p_\alpha P_{{\rm s},\alpha}\quad\mbox{subject to}
\;\;
\sum_{\alpha=1}^{n+1} p_\alpha P_{{\rm e},\alpha}\le R.
\end{equation}
Recall now that the POVMs ${\cal E}_\alpha$ are independent 
and each of them is parametrized through Eq.~(\ref{optimal-phi}) by a margin~\mbox{$r=r_\alpha$} which, moreover, satisfies the constraint~$P_{{\rm e},\alpha}\le r_\alpha$.  Therefore, Eq.~(\ref{max_probl}) can be cast as
\begin{eqnarray}
&\displaystyle
P_{\rm s}=
\max_{\{r_\alpha\}
}
\sum_{\alpha=1}^{n+1} p_\alpha P_{{\rm s},\alpha}^{W}(r_\alpha)
&\nonumber\\[.5em]
&\mbox{subject to}&\label{max-r-j} \\[.5em]
&\displaystyle
\sum_{\alpha=1}^{n+1} p_\alpha r_\alpha\! =\! R, &
\nonumber
\end{eqnarray}
where the functions $P_{{\rm s},\alpha}^W$ are defined as in Eq.~(\ref{weak}) with~\mbox{$c=c_\alpha$}. In other words, these functions give the success probability of discrimination in the subspaces~${\cal H}_\alpha$ with {\em weak} error margins $r_\alpha$.
%
The maximization of  the success probability 
translates into finding the optimal set of weak margins $\{r_\alpha\}_{\alpha=1}^{n+1}$ 
whose average, $\sum_{\alpha=1}^{n+1} p_\alpha r_\alpha$, equals a (global) margin~$R$.

Let us start by discussing the extreme cases of this scheme. On the unambiguous side, $R=0$, the only possible choice is $r_\alpha=0$ for all values of~$\alpha$, and the success probability is given by~(\ref{optUA}). At the other end point, if~$R\ge R_c=\sum_{\alpha=1}^{n+1} p_\alpha r_{c,\alpha}$, where $r_{c,\alpha}$ is the critical margin in the subspace~${\cal H}_\alpha$, given by~\eqref{critical-r} 
with $c=c_\alpha$, we immediately recover the minimum-error result~(\ref{optME}). We will refer to~$R_c$ as the global critical margin.

An explicit expression for $P_{\rm s}$ if $0 < R < R_c$ is most easily derived by starting at the unambiguous end and progressively increasing the margin $R$. For a very small error margin,
%
the Lagrange multiplier method provides the maximum. It occurs at $r_\alpha=r^{(1)}_\alpha$, where
%
%
\begin{equation}\label{lagrange}
r^{(1)}_\alpha=\frac{1-c_\alpha}{\sum_{\alpha=1}^{n+1} p_\alpha (1-c_\alpha)} R \, .
\end{equation}
%
This solution is valid only when all (partial) error margins are below their critical values,
$ r^{(1)}_\alpha\le r_{c,\alpha}$. If this inequality holds, the maximum success probability is~$P_{\rm s}=\sum_\alpha p_\alpha P_{{\rm s},\alpha}^W(r^{(1)}_\alpha)$. 
%
%
 The use of the superscript ``$(1)$'' will become clear shortly.

If we keep on increasing the global margin~$R$, it will eventually reach a value
$R=R_1$ at which the error margin 
of the first subspace ${\cal H}_1$ is saturated, namely,  where $r^{(1)}_1=r_{c,1}$. 
%
%
This is so because
the overlaps, given in Eq.~\eqref{the_c_alpha}, satisfy \mbox{$c_1 < c_2 < \hdots < c_{n+1}=1$}.
%
Hence we have
$r_1^{(1)}\! > \!r_2^{(1)} \!>\! \hdots >\! r_{n+1}^{(1)}$ and
\mbox{$r_{c,1}\!<\!r_{c,2}\!<\dots\!<\!r_{c,n+1}$}, according to~\eqref{lagrange} and~(\ref{critical-r}), respectively.
The expression for $R_1$ can be read off from Eq.~\eqref{lagrange}:
\begin{equation}
R_1=\frac{r_{c,1}}{1-c_1} \sum_{\alpha=1}^{n+1} p_\alpha (1-c_\alpha) \, .
\end{equation}
For $R>R_1$, the optimal value of the margin of subspace~${\cal H}_1$ is then frozen at the value $r_{1}=r_{c,1}$, and the remaining margins are obtained by excluding the fixed contribution of the subspace~${\cal H}_1$, i.e., by computing the maximum on the right-hand side of
%
%
\begin{eqnarray}
&\displaystyle 
P_{\rm s}-p_1 P_{{\rm s},1}^W(r_{c,1})=
\max_{\{r_\alpha\}
} \;\sum_{\alpha=2}^{n+1} p_\alpha P_{{\rm s},\alpha}^{W}(r_\alpha)
&\nonumber
\\[.5em]
&\text{subject to}& \label{r-1}
\\[.5em]
&\displaystyle
\sum_{\alpha=2}^{n+1} p_\alpha r_\alpha = R-p_1 r_{c,1}\, .
&\nonumber
\end{eqnarray}
%
The location of this maximum, which we denote by $\{r_\alpha^{(2)}\}_{\alpha=2}^{n+1}$, is formally given by~\eqref{lagrange} with $R$ replaced by $R-p_{1} r_{c,1}$ and the sum in the denominator running from~$\alpha=2$ to~$n+1$. 
In this case, we have
\begin{equation}\label{P_s at 1}
P_{\rm s}=p_1 P_{{\rm s},1}^W(r_{c,1})+\sum_{\alpha=2}^{n+1}p_\alpha
P_{{\rm s},\alpha}^W(r_{\alpha}^{(2)}) .
\end{equation}
Again, this is valid only until $R$ reaches a second saturation point $R_2$, i.e., provided $R_1<R<R_2$, and so on. Clearly, the margins $r_\alpha$ saturate in an orderly fashion as we increase $R$. 

%
%
%
%

Iterating the procedure described above, the optimal error margins in the interval $R_{\beta-1}\le R\le R_{\beta}$ (throughout the paper, Greek indexes run from $1$ to $n+1$), where $R_0\equiv0$ and $R_{n+1}\equiv R_c$, are found to be
%
%
\begin{equation}\label{lagrange k}
r^{(\beta)}_\alpha=\frac{1-c_\alpha}{\chi_\beta} \left(R-\xi_\beta \right) \, ,
\end{equation}
where
%
%
\begin{equation}\label{R-k}
R_\beta = \displaystyle\frac{r_{c,\beta} }{1-c_\beta} \chi_{\beta}+\xi_{\beta} \, , 
\end{equation}
and
\begin{equation}
\xi_\beta = \sum_{\alpha=1}^{\beta-1} p_\alpha r_{c,\alpha} \, ,
\qquad  \chi_\beta = \sum_{\alpha=\beta}^{n+1} p_\alpha (1-c_\alpha) \, .
\end{equation}
The success probability in this interval [analogous to Eq.~(\ref{P_s at 1})] is
\begin{equation}
P_{\rm s}=P^{\rm sat}_{{\rm s},\beta}+\sum_{\alpha=\beta}^{n+1}p_\alpha P_{{\rm s},\alpha}^W(r_\alpha^{(\beta)}),
\end{equation}
where
%
%
\begin{eqnarray}
P^{\rm sat}_{{\rm s},\beta} &=& \sum_{\alpha=1}^{\beta-1} p_\alpha P_{{\rm s},\alpha}(r_{c,\alpha})
\nonumber
\\[.5em]
 &=& \frac{1}{2}\sum_{\alpha=1}^{\beta-1} p_\alpha \left( 1+\sqrt{1-c_\alpha^2} \right)
 \label{ps-appendix}
\end{eqnarray}
is the contribution to the success probability of the subspaces where the error margins are frozen at their critical values.
After some algebra, we find that the success  probability can be written in a quite compact form as
%
%
\begin{equation}\label{ps-general}
P_{\rm s}\!=\! P^{\rm sat}_{{\rm s},\beta} +\left(\! \!\sqrt{R\! -\! \xi_\beta} + \sqrt{\chi_\beta} \right)^{\!2},\quad
R_{\beta-1}\!\le R\le R_{\beta}.
\end{equation}
%
Eqs.~\eqref{lagrange k} to~(\ref{ps-general}) comprise our main result.


\subsection{Strong error margin}


The concept of a strong margin for programmable machines requires a more careful formulation than that of a weak margin since, in principle, there are different conditions one can impose on the various probabilities involved. For instance, one could require the strong conditions (\ref{strong1}) and (\ref{strong2})  for \textit{every} possible pair of states fed into  the machine, that is, for every given $\{\rho_1=[\psi_1],\rho_2=[\psi_2]\}$. This approach is quickly seen to be trivial since the machine, whose performance is independent of the states, is required to satisfy the condition in a worst case scenario, in which $\ket{\psi_1}$ and $\ket{\psi_2}$ are arbitrarily close to each other. For any value of the error margin less than~$1/2$ the inconclusive probability must then approach unity, i.e.,~$Q \to1$. This implies that both $P_{\mathrm{s}}$ and 
$P_{\mathrm{e}}$ vanish. A similar argument leads to the trivial solution~$P_{\rm s}=P_{\rm e}=1/2$ if the margin is larger than or equal to~$1/2$.

The task performed by a programmable discriminator can be most naturally viewed as state labeling:  
the machine attaches the label $1$ ($2$) to the data if its state is identified, by a ``clicking'' of the operator~$E_1$~($E_2$), to be that of the qubits loaded through  program port $A$ ($C$); i.e., the state of the ports has the pattern $[\psi_1^{\otimes n}][\psi_1^{\otimes n'}][\psi_2^{\otimes n}]$ ($[\psi_1^{\otimes n}][\psi_2^{\otimes n'}][\psi_2^{\otimes n}]$).
For this task, the relevant error probabilities are $p(2|E_1)$ and $p(1|E_2)$, namely, the probability of wrongly assigning the labels 1 and 2, respectively.
 It seems, therefore, more suitable for programmable discrimination to impose the strong margin conditions $p(2|E_1)\le R$ and $p(1|E_2)\le R$.
 In terms of the average states $\sigma_1$ and $\sigma_2$ in~(\ref{average states}) these conditions~are
%
%
%
\begin{equation}\label{strongprog}
p(2|E_1)=\frac{\tr E_1 \sigma_2}{\tr E_1 \sigma_1 + \tr E_1 \sigma_2} \le R \, ,
\end{equation}
and likewise for $p(1|E_2)$.

Note that in contrast to the weak case, here the conditional probabilities are nonlinear functions of the POVM elements, and thus the maximization of the success probability under these conditions is {\em a priori} more involved. To circumvent this problem, we can use the relation~(\ref{mwms1}), which 
for programmable discrimination also holds, and reads
\begin{equation}\label{MWMS}
R^S=\frac{R^W}{P_{\rm s}(R^W)+R^W}
\end{equation}
to express the (global) weak  error margin $R^W$ in terms of the strong one $R^S$. 
%
Then, one simply uses Eqs.~\eqref{lagrange k} to~(\ref{ps-general}) to obtain the maximum success probability. 
The inversion of Eq.~(\ref{MWMS}) is somewhat lengthy but straightforward.  The difficulty arises from the fact that the success probability, Eq.~(\ref{ps-general}), is a piecewise function whose expression depends specifically on how many 
margins $r_\alpha$ have reached their critical value $r_{c,\alpha}$ for a given~$R^S$. 
Thus we need to compute the strong saturation points~$R_\beta^S$, analogous to~\eqref{R-k}, through the relation~\eqref{MWMS}.

\section{Discussion of the results}

\begin{figure}[t]
\includegraphics[scale=1.1]{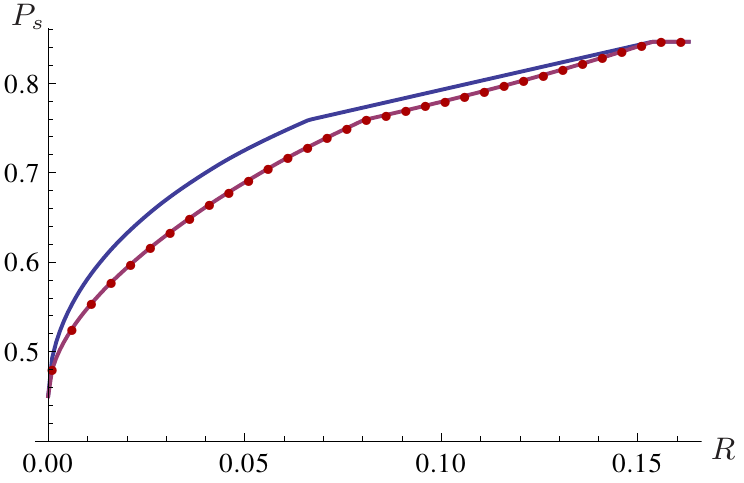}
\caption{(Color online) $P_{\rm s}$ vs $R$ for a weak (upper line) and a strong (lower line) condition, for $n=9$ and $n'=2$. The global critical margin is $R_c \simeq 0.154$. A numerical maximization of the success probability under the strong condition~(\ref{strongprog})  (points) is seen to agree with our analytical solution.}\label{fig1}
\end{figure}

\begin{figure}[t]
\includegraphics[scale=.34]{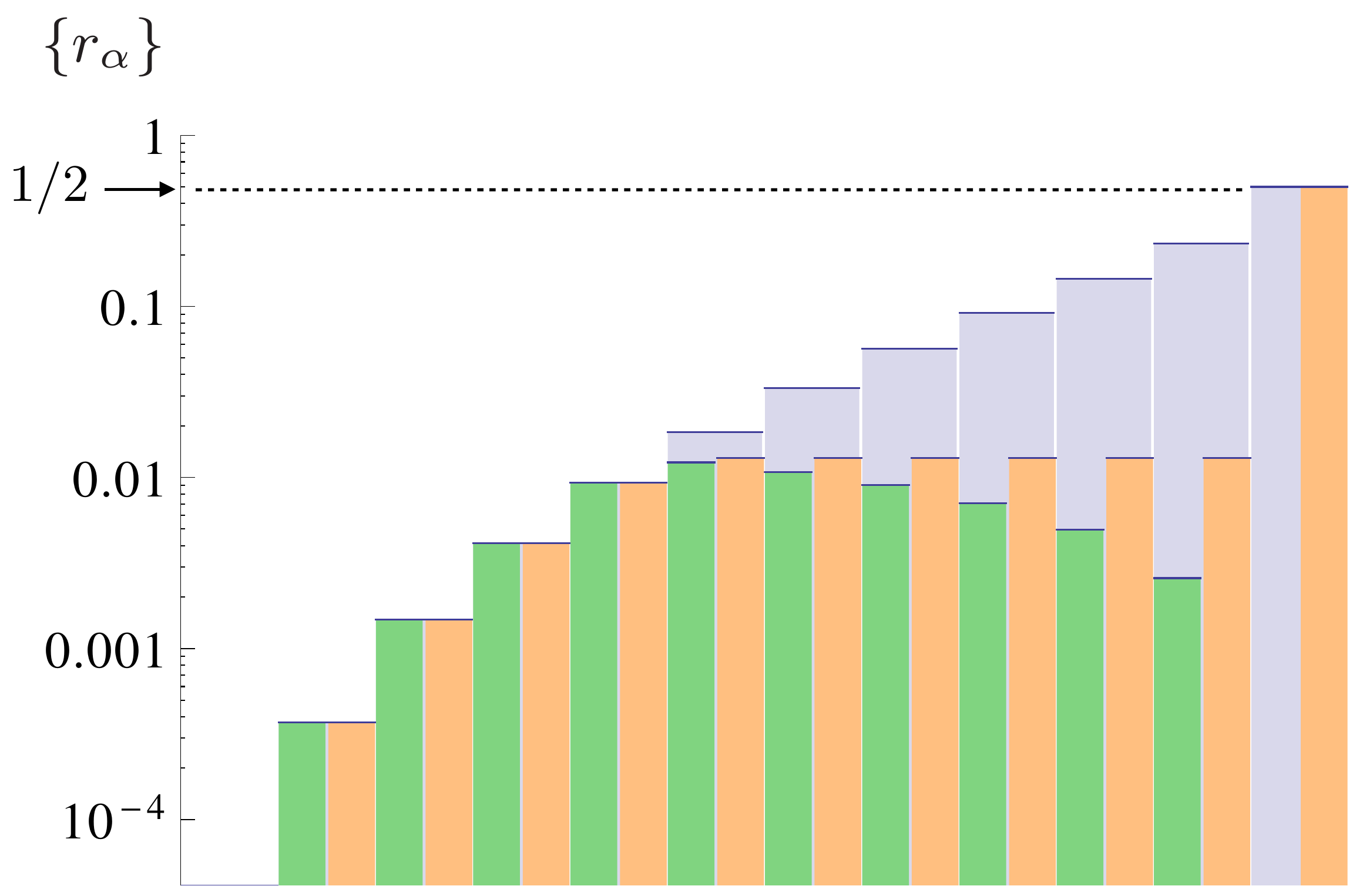}
\caption{(Color online) 
The various error margins for $n=11$, \mbox{$n'=2$} and a (global) margin~$R=0.0055$. The full heights of the wide bars in the background  (blue)  represent the values of the critical margins $r_{c,\alpha}$, 
starting from $\alpha=1$ (leftmost) up to~$\alpha=12$ (rightmost). 
For the same values of $\alpha$, each pair of narrow bars represents 
the weak margin $r_\alpha^W$ [left (green)] and the strong margin~$r_\alpha^S$ [right (orange)]. 
We note that the first five error margins have reached their critical value. The values  
for $\alpha=1$ are very small, which explains why the corresponding bars do not show up in the chart.}\label{fig2}
\end{figure}

In Fig.~\ref{fig1} we plot the maximum success probabilities for both the weak and the strong conditions
as a function of a common (global) margin $R$, for nine program and two data  copies.
We also show in Fig.~\ref{fig1} the results of a numerical optimization with the strong condition (dots), which exhibit perfect agreement with our analytical solution.
We observe that by allowing just a 5\% error margin, the success probability increases by more than~50\%.
This is just an example of a general feature of programmable discrimination with an error margin: the success probability increases sharply for small values of the error margin.

A comment about the effect of the subspace ${\cal H}_{n+1}$ 
on the shape of the plots is in order. 
This subspace contains the completely symmetric states of the whole system~$ABC$ and, hence, it is impossible to tell if the state of the data~($B$) coincides with that of one program~($A$) or that of the other~($C$);
more succinctly, $c_{n+1}=1$.
Therefore, half the number of conclusive answers will be correct and half of them will be wrong, and~$P^W_{{\rm s},n+1}=r_{n+1}$, provided~$r_{n+1}\le r_{c,n+1}=1/2$. 
Increasing the error margin simply allows for an equal increase in the success probability. 
This is reflected in the linear stretch in the upper curve 
in Fig.~\ref{fig1},
right before the (rightmost) flat plateau. For the strong condition, 
the same situation arises in the interval $R_n^S\le R\le R_c$, but the plot of the success probability is {\em not} a straight line due to the nonlinear relation~\eqref{MWMS} between the weak and the strong margin.

An alternative (though completely equivalent) way to compute the maximum success probability with a strong margin is based on the observation that the POVMs ${\cal E}_\alpha$ are also fully determined by strong margins, $r_\alpha^S$, through Eq.~(\ref{phi-strong}), with the exception of~${\cal E}_{n+1}$, for which~$c=c_{n+1}=1$ [giving rise to an ambiguity, as discussed after Eq.~(\ref{phi-strong})].
In this approach, the success probability becomes a convex combination  of $P^S_{{\rm s},\alpha}(r_\alpha^S)$, as in~(\ref{weakj}), where these functions are given in~(\ref{strong}) with~$c=c_\alpha$.
%
The optimal set~$\{r^{S\,(\beta)}_{\alpha}\}$ can be readily obtained from the weak margins in Eq.~\eqref{lagrange k} using the relation \eqref{mwms1}. 
%
The strategy in the last subspace~${\cal H}_{n+1}$ can be easily seen to consist~in abstention with a certain probability, and a random choice of the labels~1 \mbox{and~2~otherwise.}

The bar chart in Fig.~\ref{fig2} represents an optimal strategy in terms of the corresponding weak and strong error margins. For this example we have chosen~$11$ program and two data copies. For illustration purposes, the (global) margin is set to a low value of~$0.0055$. The wide vertical bars in the background depict the critical margins $r_{c,\alpha}$. There are~12 of them, displayed in increasing order of~$\alpha$ (the first one is not visible because of the small value of~$r_{c,1}$). On their left (right) halves, a narrow green (orange) bar depicts the optimal weak (strong) margin $r_\alpha^W$ ($r_\alpha^S$) (we attach the subscripts~$W$ and~$S$ through the rest of the paper to avoid confusion). We note that the first five margins ($\alpha\le5$) have reached their critical value. For $\alpha>5$, the weak margins decrease monotonically according to Eq.~\eqref{lagrange k}. For the last one, we have~$r^W_{n+1}=r^W_{12}=0$, which holds for any value of $R$, provided $R\le R_n$. This must be so, since we recall that the projections of $\sigma_1$ and $\sigma_2$ onto the subspace with maximum angular momentum are indistinguishable. Clearly,  allowing  for $r_{n+1}^W>0$ while there is still room for the other margins to increase cannot be optimal.

Also noticeable in Fig.~\ref{fig2} is that the set of strong margins that have not reached their critical  value $r_{c,\alpha}$  has a flat profile (this does not apply to $r^S_{n+1}$ that is always frozen to its critical value of~$1/2$). 
To provide an explanation for this, we write the equality in~Eq.~\eqref{strongprog}, which is attained if $R\le R_c$,  as~$R P_{\rm s} - (1-R) P_{\rm e}=0$, using once again the symmetry of the problem. We next write the success and error probabilities as a convex sum over~$\alpha$ and use the equality in the strong conditions~(\ref{strong1}) and~(\ref{strong2}) for each subspace ${\cal H}_\alpha$ to express~$P_{{\rm e},\alpha}^S$ in terms of~$P_{{\rm s},\alpha}^S$. We obtain the strong condition 
%
\begin{equation}\label{lagrange-strong}
\sum_\alpha p_\alpha P_{{\rm s},\alpha}^S (r_\alpha^S) \left[ R  - (1-R) \frac{r_\alpha^S}{1-r_\alpha^S} \right] =0 .
\end{equation}
The terms in square brackets can be positive or negative depending on $r^S_\alpha$ being smaller or larger than $R$, both of which are possible.
So, at face value, this equation cannot explain the flat profile of $r_\alpha^S$ and more work is needed.
Next, we use the Lagrange multiplier method to maximize $P_{\rm s}=\sum_\alpha p_\alpha P_{{\rm s},\alpha}^S (r_\alpha^S)$ and note that the dependence of~$P_{{\rm s},\alpha}^S$ on~$\alpha$  (i.e., the term $1-c_\alpha$) factorizes, as can be checked from Eq.~(\ref{strong}). Without further calculation, we can anticipate that the optimal margins will be determined by $n+1$ equations of the form $p_\alpha(1-c_\alpha)f(r^S_\alpha)=0$, where $f$ can be a function only of $R$, the Lagrange multiplier and the number of margins below their critical value. Hence, all the (unfrozen) margins will have the same optimal value.
%
%
For $\beta=1$ (no frozen margins) we have the simple solution $r_\alpha^{S,(1)}=R$ for all~$\alpha$, and the corresponding success probability is
\begin{equation}
P_{{\rm s}}=\left(\frac{\sqrt{1-R}}{\sqrt{R}-\sqrt{1-R}}\right)^{\!\!2}  \frac{n n'}{(n+1)(n'+2)} 
\end{equation}
for a sufficiently small strong margin $R$.


%
%

\section{Conclusions}

In this paper, we have provided two generalizations of programmable state discrimination that enable control on the rate with which errors inevitably arise because of the very principles of quantum mechanics. In the first, a margin is set on the average error probability of mislabeling the input data states (weak condition). In the second, a more stringent condition is required that, for each label, 
the probability of it being wrongly assigned is within a given margin (strong condition). Generically, in both cases, the discrimination protocol may result sometimes  in an inconclusive  outcome (i.e., in being unable to assign a label to the data).   We have shown that there is a one-to-one correspondence between these  two margins, so that weak and strong conditions turn out to be the same 
if their margins are related by a simple equation.
%
%
%
%
These generalizations extend the range of applicability of programmable discriminators  to 
scenarios where  
some rate of errors and some rate of inconclusive outcomes are both affordable; or more specifically, 
to situations where  a trade-off between these two rates is acceptable, which departs  from the standard  unambiguous (zero error) and  minimum-error (zero abstention) discrimination scenarios. 

Our results include the analytical expression of the success probability for the optimal programmable device as a function of both weak and strong error margins, as well as the characterization of the POVM that specifies such optimal device.
From the analysis of these results, we conclude that small error margins can significantly boost the success probability; i.e.,  a small departure from the unambiguous scheme can translate into an important increase of the success rate while still having very reliable results (very low error rate). We provide an example of this, where a mere error margin value of~$5\%$ adds about $50\%$ to the success probability.
%

A future extension of this work is, e.g.,  the asymptotic analysis of  programmable discrimination with an error margin, when the data and/or program ports are fed with an asymptotically large number of copies. Also relevant is the analysis of programmable discriminators when the measurement is restricted to those compatible with a machine learning scenario. These devices require only classical memory to store the information about the state of the programs, and use it in a later test stage to fix the measurement on the unknown data. They can be reused an arbitrary number of times without reloading the program ports~\cite{us-rep}.



\section*{Acknowledgments}

We acknowledge financial support from ERDF: European Regional Development Fund. This research was supported by the Generalitat de
Catalunya CIRIT, Contract No. 2009SGR-0985, by the Spanish MINECO, through Contract No. FIS2008-01236 and FPI Grant No. BES-2009-028117 (G.S.), and by the National Research Foundation~\& Ministry of Education, Singapore. Early discussions with Janos Bergou are gratefully acknowledged.

\end{document}